\documentclass[10pt]{article}
\usepackage[OE]{express}

\usepackage{multirow}
\usepackage{bm}

\newcommand*{\dif}{\mathop{}\!\mathrm{d}}
\newcommand{\tabincell}[2]{\begin{tabular}{@{}#1@{}}#2\end{tabular}}

\begin{document}
\title{Point-ahead demonstration of a transmitting antenna for satellite quantum communication}

\author{Xuan Han,\authormark{1,2,$\dagger$} Hai-Lin Yong,\authormark{1,2,$\dagger$} Ping Xu,\authormark{1,2} Wei-Yang Wang,\authormark{1,2} Kui-Xing Yang,\authormark{1,2} Hua-Jian Xue,\authormark{1,2} Wen-Qi Cai,\authormark{1,2} Ji-Gang Ren,\authormark{1,2,*} Cheng-Zhi Peng,\authormark{1,2,**} and Jian-Wei Pan\authormark{1,2}}

\address{\authormark{1}Department of Modern Physics and Hefei National Laboratory for Physical Sciences at the Microscale, University of Science and Technology of China, Hefei 230026, China\\
\authormark{2} Chinese Academy of Sciences (CAS) Center for Excellence and Synergetic Innovation Center in Quantum Information and Quantum Physics, University of Science and Technology of China, Shanghai 201315, China
}

\email{\authormark{*}jgren@ustc.edu.cn}
\email{\authormark{**}pcz@ustc.edu.cn} 
\email{\authormark{$\dagger$}These authors contributed equally to this work.} 



\begin{abstract}
A low-divergence beam is an essential prerequisite for a high-efficiency long-distance optical link, particularly for satellite-based quantum communication. A point-ahead angle, caused by satellite motion, is always several times larger than the divergence angle of the signal beam. We design a novel transmitting antenna with a point-ahead function, and provide an easy-to-perform calibration method with an accuracy better than 0.2 $\mu rad$. Subsequently, our antenna establishes an uplink to the quantum satellite, Micius, with a link loss of 41--52 dB over a distance of 500--1,400 km. The results clearly confirm the validity of our model, and provide the ability to conduct quantum communications. Our approach can be adopted in various free space optical communication systems between moving platforms.
\end{abstract}



\section{Introduction}

Quantum communication enables unconditional secure communications and fundamental tests of quantum physics \cite{Bennett:BB84:1984,Bell_Ineq_64,bennett:1993:tele,Rarity:Satellite:2002, Zeilinger_Space_04}.
In the past decades, quantum communication networks via fiber channels have been constructed and widely used in the fields of national defense, finance, etc. \cite{Peev:SECOQC:2009,Chen:Metropolitan:2010,Sasaki:TokyoQKD:2011}.
Satellite-based quantum key distribution, as a solution to global quantum communication, has become a hot research topic \cite{Satellite:Canada:NJP2013,Vallone:Single:2015,Bourgoin:Pickup:2015,Gunthner:Satellite:2017,Japan:satellite:2017}.
In 2016, a quantum science satellite called Micius was launched into a low Earth orbit (LEO). Satellite-to-ground quantum key distribution, entanglement distribution and ground-to-satellite teleportation experiments have been realized \cite{Liao:SatQKD:2017,Yin:SatEPR:2017,Ren:SatTele:2017}.
A recent study of free space quantum communication during daytime \cite{Liao:Daylight:2017} has further promoted more efficient operation of a ``quantum satellite constellation''.
Future global quantum communication networks based on constellation will require the establishment of various optical links with low channel losses, including satellite to ground, ground to satellite, and satellite to satellite links.

Many theorical and technical studies of satellite quantum communications have been performed, including link loss analysis, noise estimation and polarization basis compensation \cite{bonato:feasibility:2009,Meyer:50uplink:2011,Yin:single:2013,Zhang:Polarization:2014}. Channel loss is a factor that is more important for quantum communications than it is for classical communications since quantum signals are weaker and cannot be amplified.
Optical pointing and tracking techniques are critical tools for reducing link losses.
The influence of the relative motion between the transmitter and the receiver must thus be considered, because of the finite speed of light and the small divergence angle of the transmitting beam. Consequently, the point-ahead technique is indispensable for such quantum communication terminals.
In previous works\cite{Shapiro:point:1975,Basu:pointahead:2009}, the discussions of point-ahead focus on complex coordinate transformations.
However, discussions and detailed data analyses on actual implementation are sparse.

We construct a ground-to-satellite quantum communication link with two-stage bidirectional tracking with the PID algorithm. 
An optical uplink within 52 dB is established and well maintained with tracking techniques, including the point-ahead method.
To reduce the influence from atmospheric turbulence, our transmitter is located at an observatory ground station on the Tibetan Plateau.
We analyze the reference frames of the tracking cameras on our telescope, and calibrate the system utilizing stars' motions, and obtain a precise point-ahead model superior to 0.2 $\mu rad$.
When the LEO satellite Micius passes overhead, photons with a calibrated power sent from the ground transmitter are used to measure the link loss.
After arriving at the satellite, some of the photons are then coupled into single photon detectors,
and the photon events are logged into a time-digital-convertor (TDC) device.
From a series of measurements, we can confirm that the error from our point-ahead model is less than 0.2 $\mu rad$, and less than 3 $\mu rad$ tracking error.
As a result, the total uplink channel loss is about 41--52 dB along a 500--1400 km channel for a common cloudless day.

The rest of this paper is arranged as follows: In Section 2, we describe the point-ahead scheme for the transmitter, starting from the an analysis of the baseline requirements through optical design and implementation. Additionally, the calibration method of the point-ahead angle is also explained. In Section 3, the measurement procedure and data evaluation are provided. Measurements on a real ground-to-satellite link are conducted to evaluate the effect of our model. Finally, the results are discussed and summarized in Section 4.

\section{Requirement, implementation and calibration}
\subsection{Requirement analysis of the point-ahead model}
In satellite quantum communication, if strictly locking the beacon light which is coaxial with the signal light for reciprocity-tracking to the initial set-point (closed-loop points), the transmitting beam will be late for the receiver due to the non-ignorable relative motion. In a narrow-beam transmission scenario, it is necessary to add an offset angle in the transmitter to make sure the transmitting beam strikes the receiver accurately.

\begin{figure}[ht!]
\centering\includegraphics[width=5cm]{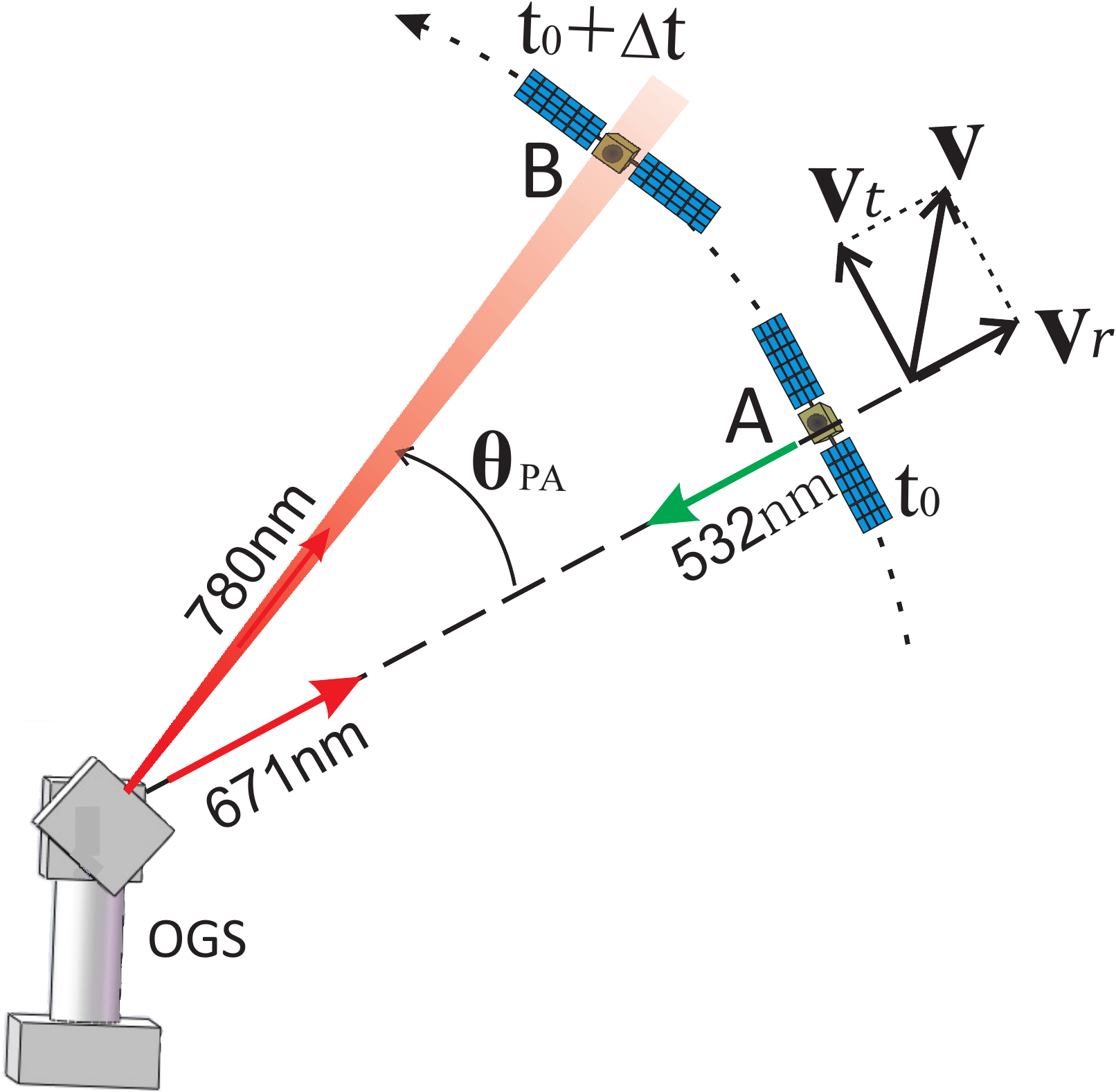}
\caption{\textbf{Configuration of the quantum communication uplink.}
The satellite (Micius) sends a green beacon of light (532 nm) at location A at time $t_0$. The optical ground station (OGS) sends a 671 nm red beacon of light, and tracks the green beacon from the satellite with a small point-ahead angle ($\bm \theta_{PA}$) offset to pre-point the next location. A 780 nm quantum signal beam sent from the OGS arrives the satellite at location B at time $t_0+\Delta t$. The relative velocity $\bm{v}$ between the satellite and the OGS can be decomposed into a radial velocity $\bm v_r$ and a tangential velocity $\bm v_t$. The point-ahead angle is proportional to tangential velocity as $\bm \theta_{PA}=2\bm v_t /c$.}
\label{FIG:pa}
\end{figure}

As shown in Fig. \ref{FIG:pa}, the relative velocity between the two terminals $\bm v$ can be decomposed into two parts: a radial velocity $\bm v_r$ and a tangential velocity $\bm v_t$. The radial velocity $v_r$ introduces the Doppler effect, which leads to a signal frequency shift at the receiver and the transmitter. The Doppler shift can be compensated by time synchronization or frequency sweep method \cite{Ho:synchronization:2009}. The shift of spectral frequency is less than 20 $pm$ and it should be considered in ultra-narrowband filtering systems.

The tangential velocity $\bm v_t$ is the origin of the need for point-ahead. We get the point-ahead angle $\bm \theta_{PA}=2\bm v_t /c$, where $\bm c$ is the speed of light. Because the velocity of the low-orbit satellites is close to the first cosmic velocity, we can determine the tangential component $\left|\bm v_t  \right|\leq 7.9 km/s$. Therefore, the point-ahead angle is $\left|\bm \theta_{PA}\right|\leq 53 \mu rad$. The maximum point-ahead angle appears when the satellite moves overhead. For deep space communication via  inter-satellite links, the relative speed between the terminals is always larger, and the maximum point-ahead angle may be more than 100 $\mu rad$. To reduce the geometrical loss over long distance link, we use a transmitter antenna with a narrow full divergence of 14 $\mu rad$. Therefore, the point-ahead influence of the relative motion must be considered to make the transmitting beam covering the receiving telescope on the satellite. The self-rotation of earth should also be considered to calculate the accurate point-ahead angles. Consider a geostationary satellite, the angles would be about 21 $\mu rad$ \cite{Shapiro:point:1975}. However, the effect is usually smaller than 0.3 $\mu rad$ for links of ground to LEO satellites. For simplification, the small bias due to coordinate systems transformation can be ignored in our demonstration.

As mentioned above, the expression of point-ahead angle is very simple, depending only on the relative tangential velocity $\bm v_t$. However, it is a practically complicated problem to implement a precise point-ahead technique in a typical system, which involves the reference transformation and the calibration errors in the terminal devices. Next, we introduce a specific transmitting antenna for sending quantum signal photons to the quantum science satellite Micius to achieve a low loss uplink with high stability.

\subsection{Implementation of a tracking system for a transmitting antenna}

In an uplink, the beam sent from the ground passes through the turbulent atmosphere at the beginning, which is utterly different from a downlink. The turbulent atmosphere degrades the wavefront and widens the divergence angle of the transmitting beam. To lower the degradation effects, we choose the Ngari Observatory ($32^\circ 19' 33.07''$ N, $80^\circ01'34.18''$ E, altitude of 5,047 m) as the OGS for excellent astronomical seeing. This site also provides a high atmospheric transmittance and a high proportion of sunny days.

\begin{figure}[ht!]
\centering\includegraphics[width=5cm]{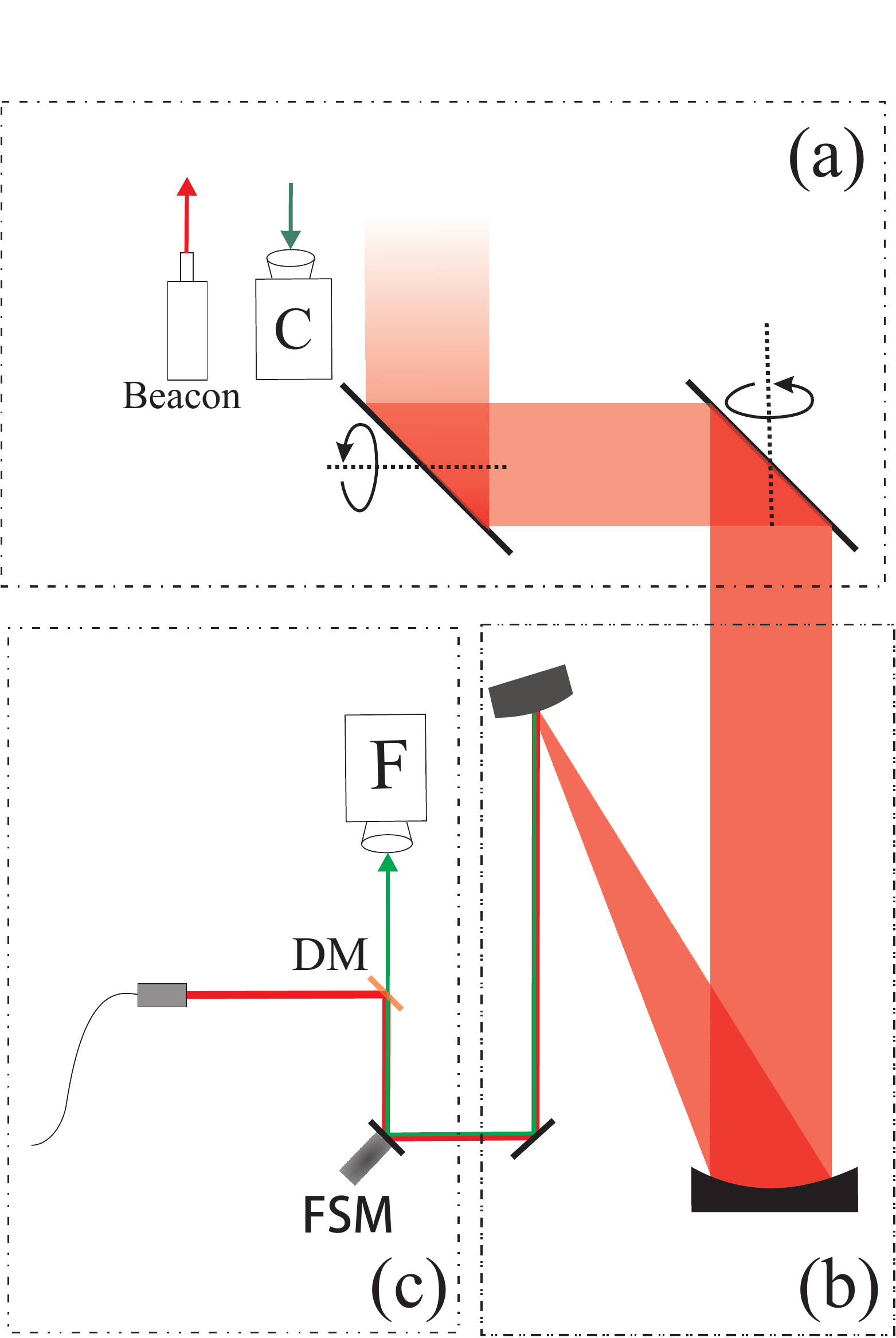}
\caption{\textbf{The quantum transmitting antenna in the Ngari Observatory.}
The transmitting antenna comprises a scanning head (a), a telescope (b) and a following optical module (c). The coarse camera mounted on the scanning head and the two-axis gimbal constitute a coarse tracking closed-loop. The beacon on the scanning head is used for satellite tracking. The telescope is a double off-axis reflective design 130 mm in diameter. The divergence of the 780 nm wavelength quantum signal beam is reduced to 14 $\mu rad$ at the output of the telescope. In the following optical module, the green beacon from the satellite passes the dichroic mirror (DM) and detected by the fine camera. The fast steering mirror (FSM) locks the beam spot to the fine-tracking point with about 1-kHz-repetition frequency. The fine tracking point is shifted to realize the point-ahead function.}
\label{FIG:antenna}
\end{figure}

As shown in Fig. \ref{FIG:antenna}, the transmitting antenna comprises a telescope with a following optical module for quantum communication and an optical scanning head. The two-dimensional scanning head can point the output beam of the telescope to any direction above the horizon.
On the scanning head, there is a coarse camera for imaging the beacon from the satellite.
In the coarse tracking, the drift between the centroid location and the closed-loop point is calculated for every frame.
The gimbal is then controlled to minimize the drift.
The error is about 10 $\mu rad$ for ground to LEO satellite tracking, with a 10-Hz acquisition and a 3-Hz feedback. 

A telescope 130 mm in diameter is used to magnify the beam to approximately 100 mm and reduce the divergence angle to approximately 14 $\mu rad$. A double off-axis reflective structure telescope is tailored to avoid optical occlusion and chromatic aberration.
In the following optical module, a 1-kHz-frame-rate high speed camera and a piezo-electric fast steering mirror (FSM) constitute a fine-tracking closed loop.

\begin{table}[ht!]
\centering
\caption{\textbf{Summary of the performances of the tracking system.}}
\label{TAB:antenna}
\begin{tabular}{|p{2.5cm}|p{3cm}|p{5cm}|}
\hline
\multicolumn{2}{|c|}{Components} & {Descriptions} \\
\hline
\multirow{4}{2cm}{Telescope}
&Type & Double off-axis reflective\\
\cline{2-3}
&Optical diameter & 130 mm \\
\cline{2-3}
&Optical magnification & 20 $\times$\\
\cline{2-3}
&Divergence for quantum signals & \tabincell{l}{Local test: 14 $\mu rad$ \\ Out of atmosphere: 15--30  $\mu rad$}\\ 
\hline
\multirow{2}{2cm}{Coarse tracking mechanism}
&Type &  \tabincell{l}{Two-axis gimbal\\(azimuth and elevation scanning)}\\
\cline{2-3}
&Tracking range& \tabincell{l}{Azimuth:$-270^\circ\sim+270^\circ$ \\ Elevation:$-5^\circ\sim+95^\circ$} \\ 
\hline
\multirow{4}{2cm}{Coarse camera}
&Type&CMOS camera, GigE interface\\ 
\cline{2-3}
&Model& TXG50 from Baumer\\ 
\cline{2-3}
&Field of view	& $\sim 1.2 ^\circ \times 1 ^\circ$ \\ 
\cline{2-3}
&Pixels \& frame rates& 2448 $\times$ 2050 pixels \& 10 Hz\\ 
\hline
\multirow{3}{2cm}{Fine tracking mechanism}
&Type & Piezo-electric fast steering mirror\\
\cline{2-3}
&Model& S-330.4SL from Physik Instrumente \\
\cline{2-3}
&Tracking range& 250 $\mu rad$  \\ 
\hline
\multirow{4}{2cm}{Fine camera}
&Type&CMOS camera, CamLink interface\\ 
\cline{2-3}
&Model& HXC40 from Baumer\\ 
\cline{2-3}
&Field of view	& 414 $\times$ 414 $\mu rad$ \\ 
\cline{2-3}
&Pixels \& frame rates&256 $\times$ 256 pixels \& 1000 Hz\\ 
\hline
\multirow{3}{2cm}{Beacon laser at ground}
&Wavelength&671 nm\\ 
\cline{2-3}
&Optical Power&2 W\\ 
\cline{2-3}
&Divergence&1.2 mrad\\ 
\hline
\multirow{3}{2cm}{Beacon laser at satellite}
&Wavelength&532 nm\\ 
\cline{2-3}
&Optical Power&160 mW\\ 
\cline{2-3}
&Divergence&1.25 mrad\\ 
\hline
\multirow{3}{2cm}{Telescope at satellite}
&Type&Cassegrain\\ 
\cline{2-3}
&Diameter&300 mm\\ 
\cline{2-3}
&Tracking Error&1.2 $\mu rad$\\
\hline
\end{tabular}
\end{table}

In addition, the scanning head is equipped with a red beacon at 671 nm wavelength with a power of 2 Watt is also equipped on the scanning head, which is adjusted parallel to the quantum signal. The divergence angle of the beacon is 1.2 mrad, which is 20 times larger than the point-ahead angle for LEO uplink. For this reason, the satellite is in the coverage of the ground beacon without point-ahead.

Specific features and parameters of the transmitting antenna and the satellite can be found in Table \ref{TAB:antenna}. The pointing requirements of the satellite receiver is also critical as it is on the ground, some parameters are also shown in Table \ref{TAB:antenna}. The divergence angle of the quantum signal is degraded to 15--30 $\mu rad$ out of atmosphere, due to air fluctuation. When the scanning head points to lower elevation direction, the quantum signal beam is observed with a larger divergence after passing through a longer atmospheric path.

\subsection{Calibration of the point-ahead calculation model based on stars}
In a practical point-ahead scheme, we only modify the closed-loop point of the fine tracking in the software without further components. In order to get the relationship between the closed-loop point of fine tracking and the orbit of satellite, we need following steps. First, the point-ahead angle changing over time according to the satellite movement is given. Second, the correlation between the offset of this closed-loop point and the structure of the transmitting antenna is presented. Finally, the transformation from coarse camera and fine camera is explained. This model can be calibrated by stars.

Given the predicted two-line elements (TLE) data of the satellite and the GPS coordinates of OGS, we can easily get the time-varying tracking parameter sequence, including azimuth-elevation coordinates of $\alpha(t)$, $\beta(t)$ and the link distance $l(t)$.
The point-ahead angle $\theta_{PA}$ for scanning head can be  decomposed to azimuth component $\alpha_{PA}$ and elevation component $\beta_{PA}$ respectively:
\begin{eqnarray}
\alpha_{PA}(t)=\dfrac{\dif{\alpha}(t)}{\dif{t}}\times \dfrac{2 l(t)}{c} \approx \dfrac{2[\alpha(t+\Delta t)-\alpha(t)]\times l(t)}{c\Delta t}\\
\beta_{PA}(t)=\dfrac{\dif{\beta}(t)}{\dif{t}}\times\dfrac{2 l(t)}{c}  \approx \dfrac{2[\beta(t+\Delta t)-\beta(t)]\times l(t)}{c\Delta t}
\end{eqnarray}

Where $\alpha(t)$ and $\beta(t)$ are the azimuth angle and the elevation angle of the transmitter at time $t$. In the calculation, $\Delta t$ can be chosen as 1 second for simplicity. For the sake of simplicity, the time-varying expression $(t)$ is omitted in the following equations.

\begin{figure}[ht!]
\centering\includegraphics[width=9cm]{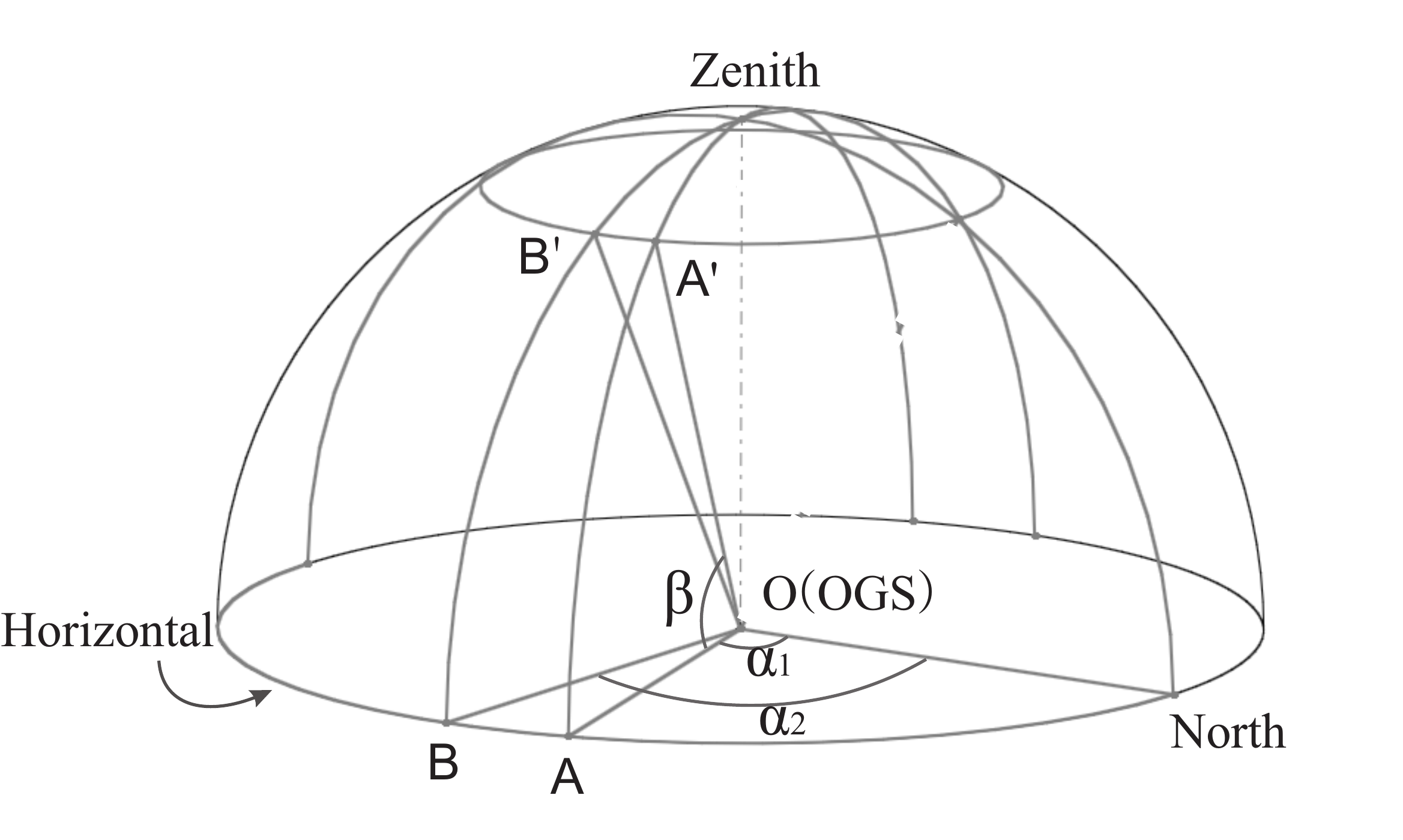}
\caption{\textbf{Actual spatial angular variation calculation from azimuth and elevation angles.}
For an azimuth-elevation telescope, the actual angle changes as the same as the angular variation in the elevation axis. However, we have $\sphericalangle AOB = \alpha_2-\alpha_1$ and $\sphericalangle A'OB'=\sphericalangle AOB \times \cos \beta = (\alpha_2-\alpha_1)\times \cos \beta$. This means, for the azimuth axis, the changed actual angle is proportional to the cosine of the elevation.}
\label{FIG:cosine}
\end{figure}

Angle transformation between scanning head and coarse camera is shown in  Fig. \ref{FIG:cosine}. The angular variation at the elevation axis is the same with the change of the elevation angle. Conversely, the angular variation at the azimuth axis is the product of the change of the azimuth angle and the cosine of the elevation angle. Further, considering the inverted image formation, we have:

\begin{equation}
\omega_{cx}=-\alpha_{PA}\times \cos \beta
\end{equation}
\begin{equation}
\omega_{cy}=-\beta_{PA}
\end{equation}

Where $\omega_{cx}$, $\omega_{cy}$ represent the point-ahead angles in X and Y directions in the coarse camera. We define $\omega_c$ as their vector sum. $\omega_f$, $\omega_{fx}$ and $\omega_{fy}$ are similarly for fine camera. We define direction angles $\theta_{c(f)}$ (0 -- 360$^\circ$, 0$^\circ$ is the positive direction on x-axis,  the angle increases in counter-clockwise direction). Then, the angle $\omega_{c(f)}$ and direction $\theta_{c(f)}$ satisfy the following relations:

\begin{equation}
\omega_{c(f)}=\sqrt{\omega^2_{c(f)x}+\omega^2_{c(f)y}}
\label{EQU:begin}
\end{equation}
\begin{equation}
\tan⁡ \theta_{c(f)}=\dfrac{\omega_{c(f)y}}{\omega_{c(f)x}}
\end{equation}

\begin{figure}[ht!]
\centering\includegraphics[width=13.5 cm]{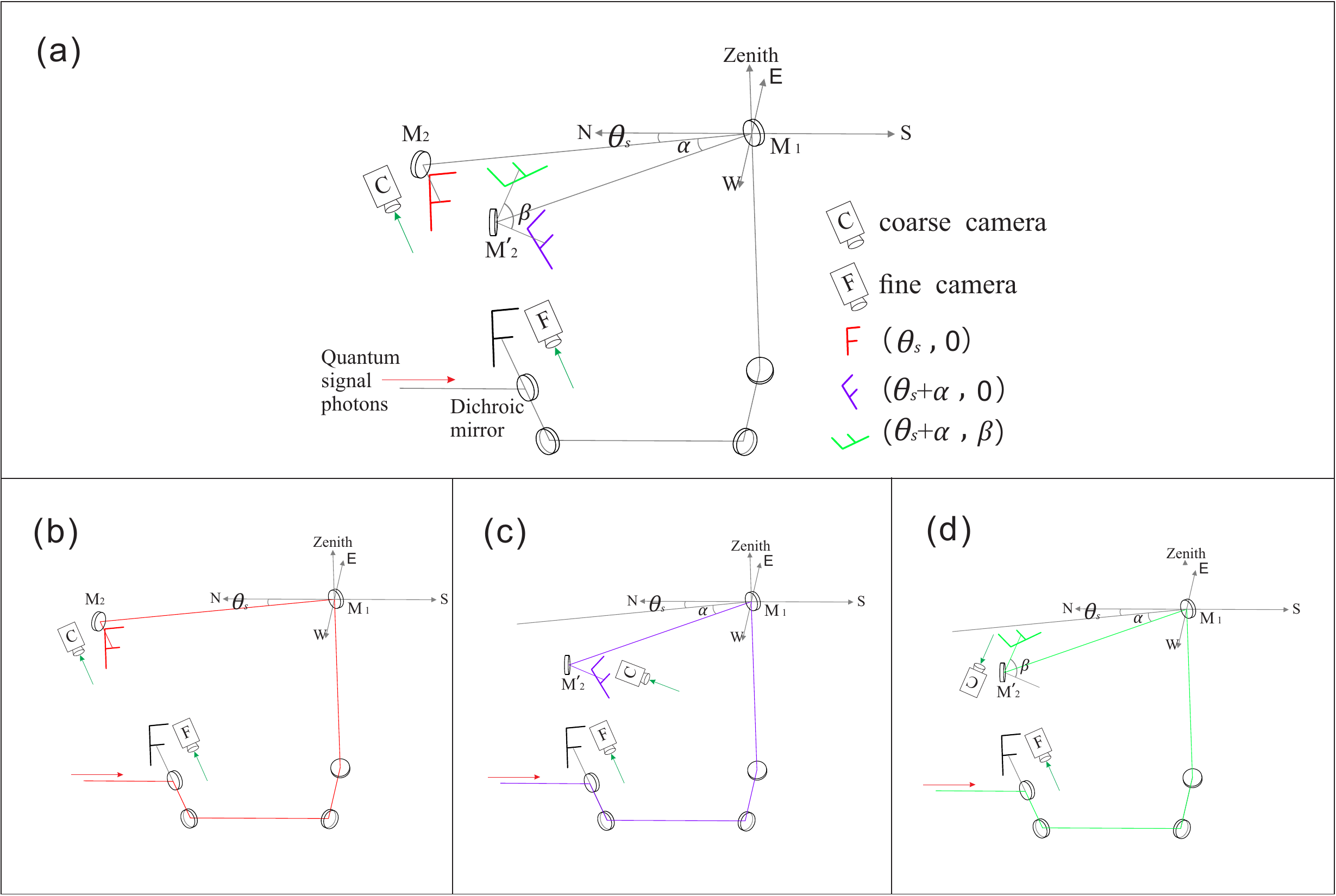}
\caption{\textbf{Image rotation between fine camera and coarse camera.}
$M1$ and $M2$ ($M2'$) are the two mirrors in the scanning head of the antenna, which drive the beam to arbitrary direction ($\alpha,\beta$) above the horizon. When the angle is ($\theta_s,0^\circ$), the direction of the beam hitting the fine camera is parallel to the one hitting the coarse camera, and no rotation exists. Here, the azimuth angle $\theta_s$ depends on what direction the antenna built on the ground. When the azimuth and elevation angles increase, an additional rotation of the same angle will appears at coarse camera in the anticlockwise direction $\Delta (\theta_c-\theta_f)=\Delta \alpha+\Delta \beta$. The image rotation between the coarse camera and the fine camera is $\alpha+\beta-\theta_s$, when the antenna points to an arbitrary direction ($\alpha,\beta$). (b), (c) and (d) represent three different directions of the antenna and (a) is their superposition.}
\label{FIG:rotation}
\end{figure}

The rotation relation between the coarse and fine camera is shown as Fig. \ref{FIG:rotation}.The normal directions of the two scanning mirrors are changing while pointing to different directions.  When the azimuth angle and the elevation angle increase, the coarse camera is rotated anticlockwise relative to the fine camera. At the same time, there is also a zero position bias $\theta_{s}$ in azimuth when installing the apparatus. For example, as shown in Fig. \ref{FIG:rotation},  no rotation exists when the coarse camera is parallel to the fine camera plane. At this time, the pointing angle of the antenna is $(\theta_{s},0^\circ)$ (Fig. \ref{FIG:rotation} (b), red $F$).When the telescope angle pointing to $(\alpha_0+\theta_{s},0^\circ)$, the relative rotation angle is $\alpha_0$ (Fig. \ref{FIG:rotation} (c), purple $F$). When the telescope angle pointing to $(\alpha_0+\theta_{s},\beta_0)$, the relative rotation angle is $\alpha_0 + \beta_0$ (Fig. \ref{FIG:rotation} (d), green $F$).
That is, for any direction $(\alpha,\beta)$, we have the following expression:

\begin{eqnarray}
\theta_{r}=\theta_c-\theta_f=\alpha+\beta-\theta_{s}
\end{eqnarray}

Taking into account the angle resolution for fine camera pixels, we set the field of view per pixel  equals to $1/k$ degree, namely the pixel resolution. Then, we have:
\begin{equation}
\omega_f=k \omega_c
\label{EQU:end}
\end{equation}
where $\omega_f$ is in pixels, $\omega_c$ is in degrees and k is a value in $\dfrac{pix}{deg}$.

\begin{figure}[ht!]
\centering\includegraphics[width=12cm]{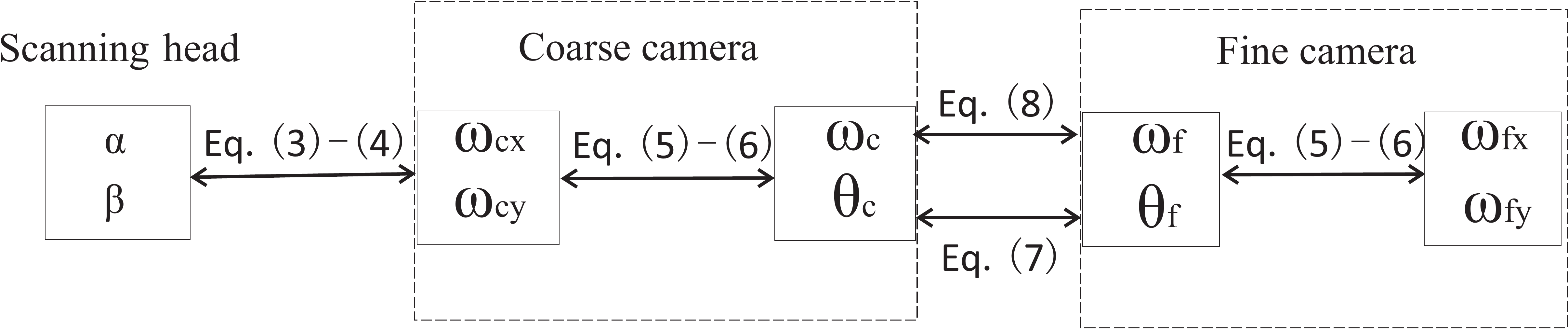}
\caption{\textbf{The relations between the scanning head angels and camera pixels.}
$\alpha$ and $\beta$ are the azimuth-elevation coordinates of the stars. 
$\omega_{c}$ and $\theta_{c}$ are the speed and the direction of the stars' movement at the scanning head.
$\omega_{f}$ and $\theta_{f}$ are the speed and the direction of the stars' movement viewed by the fine camera.
$\theta_{s}$ the azimuth angular bias due to antenna mounting. $k$ is the number of pixels equals one degree.
These parameters are in connection according to Eqs. (3)--(8).}
\label{FIG:relation}
\end{figure}

In practice, as shown in Fig. \ref{FIG:relation}, we obtain $\theta_s$ and $k$ based on tracking several bright stars for calibration.
The varying azimuth-elevation angles of the scanning head and the corresponding changing centroid coordinates of the spot in the fine camera can be easily logged with software.
In view of approximate uniform motion of the stars, linear fitting method is adopted to determine the speeds of these changing regular patterns.
After statistical analysis, we establish a point-ahead calculation model for the transmitting antenna based on Eq. (\ref{EQU:begin}) -- (\ref{EQU:end}).
In this model, the azimuth angle $\alpha$ and the elevation angle $\beta$ of the optical scanning head are connected with the numbers of fine camera pixels $\omega_{fx(y)}$ in $x$ and $y$ axes.

\begin{table}[ht!]
\centering
\caption{\textbf{Calibration of the point-ahead model with star light.} Descriptions for the parameters can be found in the caption of Fig. \ref{FIG:relation}.}
\label{TAB:calibration}
\begin{tabular}{|l|l|l|l|l|l|l|l|l|}
\hline
Parameters& $\alpha$ & $\beta$ & $\omega_{c}$ & $\theta_{c}$ &$\omega_{f}$&$\theta_{f}$& $\theta_{s}$ &$k$\\
(Units)& ($deg$) & ($deg$) & ($deg/s$) & ($deg$) & ($pix/s$) & ($deg$) & ($deg$) & \Big($\dfrac{pix}{deg}$\Big)\\
\hline
   Aries $\alpha$ & 77.5& 26.3 & $3.82 \times 10^{-3}$ & -63.95  & 41.49 & -172.79 & -5.05 & 10866 \\
\hline%
   Aries $\beta$ &80.8& 26.2 & $3.90 \times 10^{-3}$ & -63.08 & 42.51 & -174.69 & -4.58 & 10892 \\
\hline
  Ophiuchus $\alpha$ & 260.4& 37.9 &  $4.10 \times 10^{-3}$ & 59.34 &44.98 &116.22 & -4.87 & 10961 \\
\hline
 Aquila $\alpha$ & 220.3& 61.0 & $4.12 \times 10^{-3}$ & 33.75 &44.89 & 107.31 & -5.12 & 10891 \\
\hline
 Cassiopeia $\beta$ & 35.3& 47.8 & $2.14 \times 10^{-3}$ & -107.25 & 22.82 & 165.02  & -4.63 & 10682 \\
\hline
 Cepheus $\alpha$ & 81.4& 49.2 & $3.65 \times 10^{-3}$ &-72.94 & 39.42 & -151.94& -4.46 & 10800 \\
\hline
 Andromeda $\alpha$ & 5.4& 59.4 &  $1.92 \times 10^{-3}$  & -170.25 &20.37 & 120.04 & -4.88 & 10627 \\
\hline
 Perseus $\alpha$  & 43.1& 20.7 & $2.69 \times 10^{-3}$ & -64.48 & 28.72 & -131.90  & -3.57 & 10667 \\
\hline
\end{tabular}
\end{table}

In Table \ref{TAB:calibration}, the calibration parameters of eight stars are shown. The speed and the direction of the stars' movement at the scanning head and the fine camera are linear fitted from the original observation data. The azimuth angular bias $\theta_{s}$ and a ratio parameter $k$ are estimated for each star. The estimation errors for each star are analyzed with the fitting errors and error transfer formula. For the star ``Aries $\alpha$'', $\theta_{s} = -5.05 \pm 0.40$, and $k = 10866 \pm 113$. Computing the average values and estimating the final errors for the model, we have $\theta_{s} = -4.65 \pm 0.12$, and $k=10798 \pm 34$.

\begin{figure}[ht!]
\centering\includegraphics[width=12.5cm]{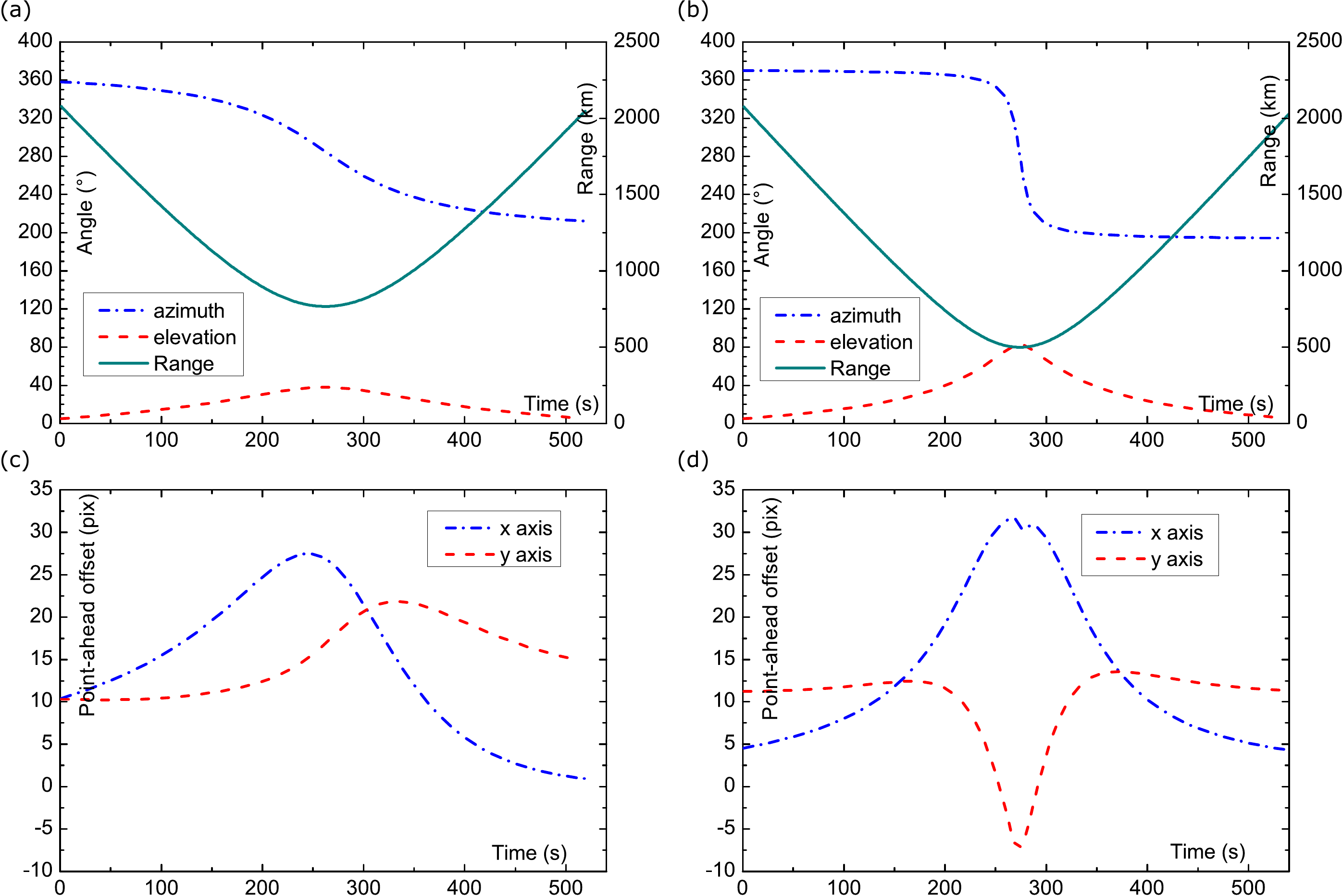}
\caption{\textbf{Two typical passages and the point-ahead angles in the fine camera.}
(a) Farther passage. The maximum elevation angle is 38 degree and the nearest location is 767 km away from the OGS.  
(c) Nearer passage. The maximum elevation angle is 83 degree and the nearest location is 499 km away from the OGS.  
(b) and (d) are the point-ahead pixels for the passage (a) and (c) in the fine camera, respectively.
}
\label{FIG:passes}
\end{figure}

\section{Results}
\subsection{Point-ahead angles and error estimation}
In a specific implementation, we first get point-ahead angles for the scanning head with Eqs. (1)--(2), and then use the transform method shown in Fig.~\ref{FIG:relation} with calibrated parameters. Later, we get point-ahead pixels in the fine camera. To give illustrations, two typical passages are shown in Figs.~\ref{FIG:passes}(a) and~\ref{FIG:passes}(c). When the passage is far from the zenith, the azimuth angle is changing smoothly. And an accelerated motion appears in the azimuth axis when the satellite passes near the zenith. The actual point-ahead pixels at the fine camera are given in Figs.~\ref{FIG:passes}(b) and~\ref{FIG:passes}(d) for the two passages in Figs.~\ref{FIG:passes}(a) and~\ref{FIG:passes}(c). At the highest elevation angle for one passage, the maximum point-ahead pixels are about 30 pixels (50 $\mu rad$).

In addition, the error from the model is also estimated under the error propagation formula. In our analysis, the inaccuracy in the parameter calibration is considered. The final errors for point-ahead pixels are estimated among 0.03--0.13 pixels in x and y axis for different passages, these are 0.05--0.2 $\mu rad$ for point-ahead angle in one axis.

\subsection{Ground-to-satellite link and efficiency scanning}

\begin{figure}[ht!]
\centering\includegraphics[width=12.5cm]{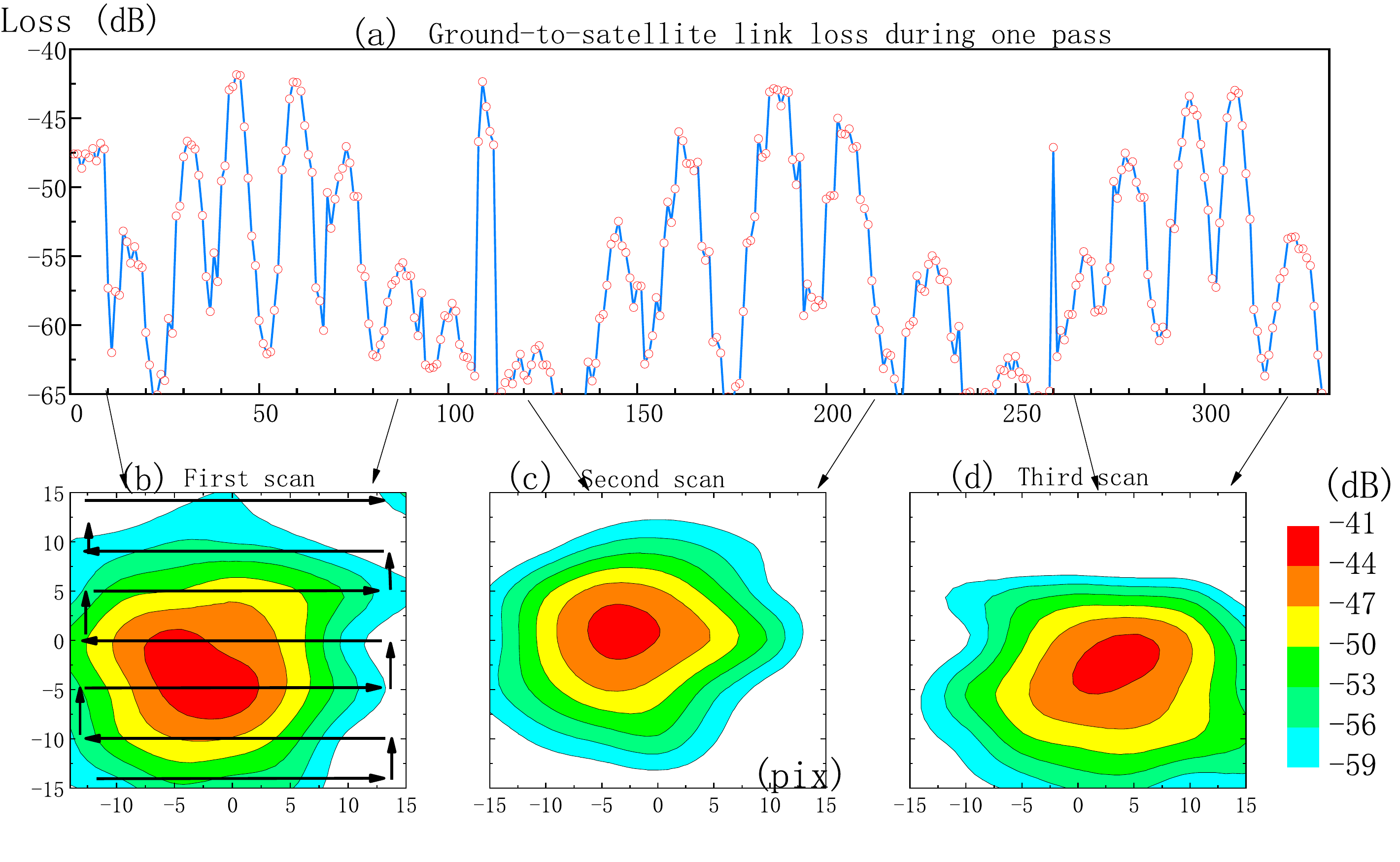}
\caption{\textbf{Three efficiency scanning test during one passage.} 
(a) The total loss for ground to satellite link. In the passage, the fine closed-loop point is the sum of the original point, the point-ahead offset and the snakelike scanning pattern.The closed-lope point will vary 3 pixels each time along x or y axis every 3 seconds. The scanning is repeated for three times during one passage, and the fitted data is shown in (b), (c) and (d).
}
\label{FIG:scan}
\end{figure}

In our experiment, a reference laser (about $2\times 10^{9}$ photons) is sent from the ground to the satellite to measure the link loss. The photon detected in the satellite is logged in the counting module\cite{Ren:SatTele:2017}. The link loss is estimated as 41--52 dB(1.5 dB for antenna transmittance, 1.5 dB for air transmittance, 30.5--40 dB for geometric loss depending on the distance, and 7.5 dB for coupling and detection efficiency) when the length of the distance is less than 1400 km.
For further examination of the point-ahead function, in one passage of the measurement, we modify the closed-loop point with a snakelike 
scanning. In this passage, the maximum elevation angle is 55$^\circ$ and the scanning is repeated three times. The time varying link loss is shown in Fig.~\ref{FIG:scan}(a), the peaks in the curve appear at a near zero offset at one axis in the snakelike scanning. Figs.~\ref{FIG:scan}(b) --~\ref{FIG:scan}(d) give the three contours for scanning test. This measurement can also serve as an evidence for good point-ahead performance.
The small drift from the zero point in the contours is mainly due to the unstable atmospheric environment and bidirectional tracking. In the end of the third scanning, the link loss is larger than 60 dB due to cloud obscuration at low elevation angle. 
Fig.~\ref{FIG:scan}(c) is scanned at the period with a high elevation angle, near 55$^\circ$ .
Although with a larger point-ahead offset, the pattern of the contour is better than others due to smaller atmospheric effects and good tracking.

\section{Conclusion}
We designed and implemented a quantum optical transmitter for ground-to-satellite uplink. Our transmitter achieves high precision tracking with two-stage tracking design. With automated offset setting on the fine-tracking point, and the point-ahead is achieved with less than 0.2 $\mu rad$ model error and 3 $\mu rad$ tracking error. Based on this work, a quantum optical link was established between ground and the quantum science satellite. The whole link attenuations were 41--52 dB. A ground-to-satellite quantum teleportation experiment was accomplished \cite{Ren:SatTele:2017}. The ground-to-satellite uplink we established is the basis of a series of quantum communication experiments, such as an uplink QKD \cite{Meyer:50uplink:2011,Satellite:Canada:NJP2013}, a decoherence test due to gravity \cite{QUEST:gravity:2017} and so on\cite{Rarity:Satellite:2002,Zeilinger_Space_04}. At the same time, the easy-to-implement and high-precision point-ahead technique we designed is also important for the free space optical communication of other moving platforms. The research and application of space quantum communication will be an important step toward a global quantum communication network.

\section*{Funding}
This work was supported by the Strategic Priority Research Program on Space Science, the Chinese Academy of Sciences, National Natural Science Foundation of China Grants U1738203 and  U1738204, and Shanghai Sailing Program.

\section*{Acknowledgments}
We thank many colleagues at Nanjing Astronomical Instruments Company Limited, the National Space Science Center, National Astronomical Observatories and Xi'an Satellite Control Centre.


\begin{thebibliography}{99}
\newcommand{\enquote}[1]{``#1''}

\bibitem{Bennett:BB84:1984}
C.~H. Bennett and G.~Brassard, \enquote{Quantum cryptography: Public key
  distribution and coin tossing,} in \emph{Proceedings of the IEEE
  International Conference on Computers, Systems and Signal Processing},
  (IEEE Press, New York, 1984), pp. 175--179.

\bibitem{Bell_Ineq_64}
J.~S. Bell, \enquote{On the {E}instein-{P}odolsky-{R}osen paradox,} Physics
  \textbf{1}, 195--200 (1964).

\bibitem{bennett:1993:tele}
C.~H. Bennett, G.~Brassard, C.~Cr{\'e}peau, R.~Jozsa, A.~Peres, and W.~K.
  Wootters, \enquote{Teleporting an unknown quantum state via dual classical
  and Einstein-Podolsky-Rosen channels,} Physical Review Letters \textbf{70},
  1895 (1993).

\bibitem{Rarity:Satellite:2002}
J.~Rarity, P.~Tapster, P.~Gorman, and P.~Knight, \enquote{Ground to satellite
  secure key exchange using quantum cryptography,} New J. Phys. \textbf{4}, 82
  (2002).

\bibitem{Zeilinger_Space_04}
R.~Kaltenbaek, M.~Aspelmeyer, T.~Jennewein, C.~Brukner, A.~Zeilinger,
  M.~Pfennigbauer, and W.~R. Leeb, \enquote{Proof-of-concept experiments for
  quantum physics in space,} in \emph{Quantum Communications and
  Quantum Imaging} (International Society for Optics and
  Photonics, 2004), vol. 5161, pp. 252--269.

\bibitem{Peev:SECOQC:2009}
M.~Peev, C.~Pacher, R.~All\'eaume, C.~Barreiro, J.~Bouda, W.~Boxleitner,
  T.~Debuisschert, E.~Diamanti, M.~Dianati, J.~F. Dynes, S.~Fasel, S.~Fossier,
  M.~F\"urst, J.-D. Gautier, O.~Gay, N.~Gisin, P.~Grangier, A.~Happe,
  Y.~Hasani, M.~Hentschel, H.~H\"ubel, G.~Humer, T.~L\"anger, M.~Legr\'e,
  R.~Lieger, J.~Lodewyck, T.~Lor\"unser, N.~L\"utkenhaus, A.~Marhold,
  T.~Matyus, O.~Maurhart, L.~Monat, S.~Nauerth, J.-B. Page, A.~Poppe,
  E.~Querasser, G.~Ribordy, S.~Robyr, L.~Salvail, A.~W. Sharpe, A.~J. Shields,
  D.~Stucki, M.~Suda, C.~Tamas, T.~Themel, R.~T. Thew, Y.~Thoma, A.~Treiber,
  P.~Trinkler, R.~Tualle-Brouri, F.~Vannel, N.~Walenta, H.~Weier,
  H.~Weinfurter, I.~Wimberger, Z.~L. Yuan, H.~Zbinden, and A.~Zeilinger,
  \enquote{The {SECOQC} quantum key distribution network in {V}ienna,} New J.
  Phys. \textbf{11}, 075001 (2009).

\bibitem{Chen:Metropolitan:2010}
T.-Y. Chen, J.~Wang, H.~Liang, W.-Y. Liu, Y.~Liu, X.~Jiang, Y.~Wang, X.~Wan,
  W.-Q. Cai, L.~Ju, L.-K. Chen, L.-J. Wang, Y.~Gao, K.~Chen, C.-Z. Peng, Z.-B.
  Chen, and J.-W. Pan, \enquote{Metropolitan all-pass and inter-city quantum
  communication network,} Opt. Express \textbf{18}, 27217--27225 (2010).

\bibitem{Sasaki:TokyoQKD:2011}
M.~Sasaki, M.~Fujiwara, H.~Ishizuka, W.~Klaus, K.~Wakui, M.~Takeoka, A.~Tanaka,
  K.~Yoshino, Y.~Nambu, S.~Takahashi, A.~Tajima, A.~Tomita, T.~Domeki,
  T.~Hasegawa, Y.~Sakai, H.~Kobayashi, T.~Asai, K.~Shimizu, T.~Tokura,
  T.~Tsurumaru, M.~Matsui, T.~Honjo, K.~Tamaki, H.~Takesue, Y.~Tokura, J.~F.
  Dynes, A.~R. Dixon, A.~W. Sharpe, Z.~L. Yuan, A.~J. Shields, S.~Uchikoga,
  M.~Legre, S.~Robyr, P.~Trinkler, L.~Monat, J.-B. Page, G.~Ribordy, A.~Poppe,
  A.~Allacher, O.~Maurhart, T.~Langer, M.~Peev, and A.~Zeilinger,
  \enquote{Field test of quantum key distribution in the {Tokyo QKD Network},}
  Opt. Express \textbf{19}, 10387--10409 (2011).

\bibitem{Satellite:Canada:NJP2013}
J.-P. Bourgoin, E.~Meyer-Scott, B.~L. Higgins, B.~Helou, C.~Erven,
  H.~H\"{u}bel, B.~Kumar, D.~Hudson, I.~D'Souza, R.~Girard, R.~Laflamme, and
  T.~Jennewein, \enquote{A comprehensive design and performance analysis of low
  {E}arth orbit satellite quantum communication,} New J. Phys. \textbf{15},
  023006 (2013).

\bibitem{Vallone:Single:2015}
G.~Vallone, D.~Bacco, D.~Dequal, S.~Gaiarin, V.~Luceri, G.~Bianco, and
  P.~Villoresi, \enquote{Experimental satellite quantum communications,} Phys.
  Rev. Lett. \textbf{115}, 040502 (2015).

\bibitem{Bourgoin:Pickup:2015}
J.-P. Bourgoin, B.~L. Higgins, N.~Gigov, C.~Holloway, C.~J. Pugh, S.~Kaiser,
  M.~Cranmer, and T.~Jennewein, \enquote{Free-space quantum key distribution to
  a moving receiver,} Opt. Express \textbf{23}, 33437--33447 (2015).

\bibitem{Gunthner:Satellite:2017}
K.~G{\"u}nthner, I.~Khan, D.~Elser, B.~Stiller, {\"O}.~Bayraktar, C.~R.
  M{\"u}ller, K.~Saucke, D.~Tr{\"o}ndle, F.~Heine, S.~Seel, P.~Greulich,
  H.~Zech, B.~G{\"u}tlich, S.~Philipp-May, C.~Marquardt, and G.~Leuchs,
  \enquote{Quantum-limited measurements of optical signals from a geostationary
  satellite,} Optica \textbf{4}, 611--616 (2017).

\bibitem{Japan:satellite:2017}
H.~Takenaka, A.~Carrasco-Casado, M.~Fujiwara, M.~Kitamura, M.~Sasaki, and
  M.~Toyoshima, \enquote{Satellite-to-ground quantum communication using a
  50-kg-class micro-satellite,} Nat. Photon. \textbf{11}, 502--508 (2017).

\bibitem{Liao:SatQKD:2017}
S.-K. Liao, W.-Q. Cai, W.-Y. Liu, L.~Zhang, Y.~Li, J.-G. Ren, J.~Yin, Q.~Shen,
  Y.~Cao, Z.-P. Li, F.-Z. Li, X.-W. Chen, L.-H. Sun, J.-J. Jia, J.-C. Wu, X.-J.
  Jiang, J.-F. Wang, Y.-M. Huang, Q.~Wang, Y.-L. Zhou, L.~Deng, T.~Xi, L.~Ma,
  T.~Hu, Q.~Zhang, Y.-A. Chen, N.-L. Liu, X.-B. Wang, Z.-C. Zhu, C.-Y. Lu,
  R.~Shu, C.-Z. Peng, J.-Y. Wang, and J.-W. Pan, \enquote{Satellite-to-ground
  quantum key distribution,} Nature \textbf{549}, 43--47 (2017).

\bibitem{Yin:SatEPR:2017}
J.~Yin, Y.~Cao, Y.-H. Li, S.-K. Liao, L.~Zhang, J.-G. Ren, W.-Q. Cai, W.-Y.
  Liu, B.~Li, H.~Dai, G.-B. Li, Q.-M. Lu, Y.-H. Gong, Y.~Xu, S.-L. Li, F.-Z.
  Li, Y.-Y. Yin, Z.-Q. Jiang, M.~Li, J.-J. Jia, G.~Ren, D.~He, Y.-L. Zhou,
  X.-X. Zhang, N.~Wang, X.~Chang, Z.-C. Zhu, N.-L. Liu, Y.-A. Chen, C.-Y. Lu,
  R.~Shu, C.-Z. Peng, J.-Y. Wang, and J.-W. Pan, \enquote{Satellite-based
  entanglement distribution over 1200 kilometers,} Science \textbf{356},
  1140--1144 (2017).

\bibitem{Ren:SatTele:2017}
J.-G. Ren, P.~Xu, H.-L. Yong, L.~Zhang, S.-K. Liao, J.~Yin, W.-Y. Liu, W.-Q.
  Cai, M.~Yang, L.~Li, X.~Kui-Xing~Yang, X.~Han, Y.-Q. Yao, J.~Li, H.-Y. Wu,
  S.~Wan, L.~Liu, D.-Q. Liu, Y.-W. Kuang, Z.-P. He, P.~Shang, C.~Guo, R.-H.
  Zheng, K.~Tian, Z.-C. Zhu, N.-L. Liu, C.-Y. Lu, R.~Shu, Y.-A. Chen, C.-Z.
  Peng, J.-Y. Wang, and J.-W. Pan, \enquote{Ground-to-satellite quantum
  teleportation,} Nature \textbf{549}, 70--73 (2017).

\bibitem{Liao:Daylight:2017}
S.-K. Liao, H.-L. Yong, C.~Liu, G.-L. Shentu, D.-D. Li, J.~Lin, H.~Dai, S.-Q.
  Zhao, B.~Li, J.-Y. Guan, W.~Chen, Y.-H. Gong, Y.~Li, Z.-H. Lin, G.-S. Pan,
  J.~Pelc, M.~M.~Fejer, W.-Z. Zhang, W.-Y. Liu, J.~Yin, J.-G. Ren, X.-B. Wang,
  Z.~Qiang, C.-Z. Peng, and J.-W. Pan, \enquote{Long-distance free-space
  quantum key distribution in daylight towards inter-satellite communication,}
  Nat. Photon. \textbf{11}, 509--513 (2017).

\bibitem{bonato:feasibility:2009}
C.~Bonato, A.~Tomaello, V.~Da~Deppo, G.~Naletto, and P.~Villoresi,
  \enquote{Feasibility of satellite quantum key distribution,} New J. Phys.
  \textbf{11}, 045017 (2009).

\bibitem{Meyer:50uplink:2011}
E.~Meyer-Scott, Z.~Yan, A.~MacDonald, J.-P. Bourgoin, H.~H{\"u}bel, and
  T.~Jennewein, \enquote{How to implement decoy-state quantum key distribution
  for a satellite uplink with 50-dB channel loss,} Phys. Rev. A \textbf{84},
  062326 (2011).

\bibitem{Yin:single:2013}
J.~Yin, Y.~Cao, S.-B. Liu, G.-S. Pan, J.-H. Wang, T.~Yang, Z.-P. Zhang, F.-M.
  Yang, Y.-A. Chen, C.-Z. Peng, and J.-W. Pan, \enquote{Experimental
  quasi-single-photon transmission from satellite to {E}arth,} Opt. Express
  \textbf{21}, 20032--20040 (2013).

\bibitem{Zhang:Polarization:2014}
M.~Zhang, L.~Zhang, J.~Wu, S.~Yang, X.~Wan, Z.~He, J.~Jia, D.~Citrin, and
  J.~Wang, \enquote{Detection and compensation of basis deviation in
  satellite-to-ground quantum communications,} Opt. Express \textbf{22},
  9871--9886 (2014).

\bibitem{Shapiro:point:1975}
J.~H. Shapiro, \enquote{Point-ahead limitation on reciprocity tracking,} J.
  Opt. Soc. Am. \textbf{65}, 65--68 (1975).

\bibitem{Basu:pointahead:2009}
S.~Basu, D.~Voelz, and D.~K. Borah, \enquote{Fade statistics of a
  ground-to-satellite optical link in the presence of lead-ahead and aperture
  mismatch,} Appl. Optics \textbf{48}, 1274--1287 (2009).

\bibitem{Ho:synchronization:2009}
C.~Ho, A.~Lamas-Linares, and C.~Kurtsiefer, \enquote{Clock synchronization by
  remote detection of correlated photon pairs,} New J. Phys. \textbf{11},
  045011 (2009).

\bibitem{QUEST:gravity:2017}
S.~{Koduru Joshi}, J.~{Pienaar}, T.~C. {Ralph}, L.~{Cacciapuoti},
  W.~{McCutcheon}, J.~{Rarity}, D.~{Giggenbach}, V.~{Makarov}, I.~{Fuentes},
  T.~{Scheidl}, E.~{Beckert}, M.~{Bourennane}, D.~E. {Bruschi}, A.~{Cabello},
  J.~{Capmany}, J.~A. {Carrasco}, A.~{Carrasco-Casado}, E.~{Diamanti},
  M.~{Duusek}, D.~{Elser}, A.~{Gulinatti}, R.~H. {Hadfield}, T.~{Jennewein},
  R.~{Kaltenbaek}, M.~A. {Krainak}, H.-K. {Lo}, C.~{Marquardt}, G.~{Milburn},
  M.~{Peev}, A.~{Poppe}, V.~{Pruneri}, R.~{Renner}, C.~{Salomon}, J.~{Skaar},
  N.~{Solomos}, M.~{Stip{\v c}evi{\'c}}, J.~P. {Torres}, M.~{Toyoshima},
  P.~{Villoresi}, I.~{Walmsley}, G.~{Weihs}, H.~{Weinfurter}, A.~{Zeilinger},
  M.~{{\.Z}ukowski}, and R.~{Ursin}, \enquote{Space {QUEST} mission proposal:
  Experimentally testing decoherence due to gravity,} arXiv:1703.08036 e-prints  (2017).

\end{thebibliography}
\end{document}